\documentclass{article}
\usepackage{amsmath,graphicx}
\usepackage{spconf}
\usepackage[pdftex]{hyperref}
\usepackage[comma,sort&compress,numbers]{natbib}
\usepackage[hang,font=small]{caption}
\usepackage{booktabs}

\graphicspath{{figures/}}

\title{Quality Assessment Methods for Perceptual Video Compression}
\name{Fan Zhang and David R. Bull \thanks{*The authors acknowledge funding from ORSAS, EPSRC and the University of Bristol.}}
\address{Department of Electrical and Electronic Engineering, University of Bristol, Bristol, BS8 1UB, UK.\\
{\{Fan.Zhang,Dave.Bull\}@bristol.ac.uk}}
\begin{document}

\maketitle
\begin{abstract}

This paper describes a quality assessment model for perceptual video compression applications (PVM), which stimulates visual masking and distortion-artefact perception using an adaptive combination of noticeable distortions and blurring artefacts. The method shows significant improvement over existing quality metrics based on the VQEG database, and provides compatibility with in-loop rate-quality optimisation for next generation video codecs due to its latency and complexity attributes. Performance comparison are validated against a range of different distortion types.

\end{abstract}

\section{Introduction}
\label{sec:intro}

With the target of improving rate-quality performance in video compression, a number of new coding approaches have been recently proposed \cite{j:Lee}. These methods focus on providing good subjective quality rather than on minimising errors between the original and coded pictures. They typically utilise texture analysis and synthesis techniques instead of (or as well as) conventional energy minimisation approaches. One of the most difficult problems associated with these perceptual video codecs is the lack of a reliable quality assessment measure to estimate subjective quality. This is needed both for in-loop rate quality optimisation (RQO) and mode decision rating and for compression performance evaluation. Quality metrics for this type of codec should therefore correlate well with subjective perception, and offer high integration flexibility and manageable computational complexity.  

It is widely known that our perceptual sensitivity to picture distortions varies with luminance level, the existence of textures, and the level of artefacts. This is often described in terms of visual masking and contrast sensitivity \cite{j:Kelly} characteristics, which have often been utilised in quality metrics such as just noticeable distortion (JND) \cite{j:ZhangXue,j:Wei} and visual signal-to-noise ratio (VSNR) \cite{j:vsnr}. This has provided a basis for many perceptual video compression algorithms including \cite{c:NdjikiNya,j:Bosch,j:Zhang}. Moreover, visual statistics and features are also employed for developing quality assessment methods \cite{j:Chikkerur}. This type of approach has been used in the structural similarity measure (SSIM) \cite{j:ssim} (using structural information), the motion tuned spatio-temporal quality assessment method (MOVIE) (using spatio-temporal information). Recently, the near-threshold and supra-threshold perception strategy has been successfully exploited in metrics such as VSNR \cite{j:vsnr} and MAD \cite{j:MAD}, where human perception is modelled as two distinct processes. Finally, Zhang and Bull proposed an artefact-based video metric (AVM) \cite{j:Zhang} for their analysis-synthesis video coding framework, which is used both in-loop for RQO as well as outside the coding framework for performance evaluation. Based on the subjective experiment results reported, this performs well for both conventionally compressed and synthesised content.

In this paper, a generic objective quality assessment method is described, inspired by AVM. This approach non-linearly combines noticeable distortion and blurring artefacts based on a complex wavelet transform decomposition and motion analysis. PVM and its lite version, PIM, provide competitive correlation performance with subjective judgements and offer efficiency and flexibility for in-loop processing. 

The organisation of the rest of the paper is given as follows: Section \ref{sec:algorithm} highlights visual masking effects and distortion artefact perception process based on the correlation analysis on the VQEG database, from which the perception-based video metrics are proposed in detail. In Section \ref{sec:results}, the performance of the proposed metrics are presented including correlation analysis, complexity and latency estimation. Finally, conclusions are provided in Section \ref{sec:conclusion}.

\section{Proposed Algorithm}
\label{sec:algorithm}

Two characteristics of the Human Visual System (HVS) are commonly exploited in objective quality metric development: visual masking and a two-stage perception strategy. Texture masking implies that greater distortion can be tolerated by the HVS in textured regions (static and dynamic) than in plain luminance areas \cite{j:Lee}. Two-stage perception theory observes that human perception tends to estimate distortion for high quality videos, while it detects artefacts in low quality cases. These two phenomena provide a basis for the proposed methods.

\begin{figure}[ht]
  \centering
  \vspace{-0.4cm}
  \centerline{\includegraphics[width=9cm]{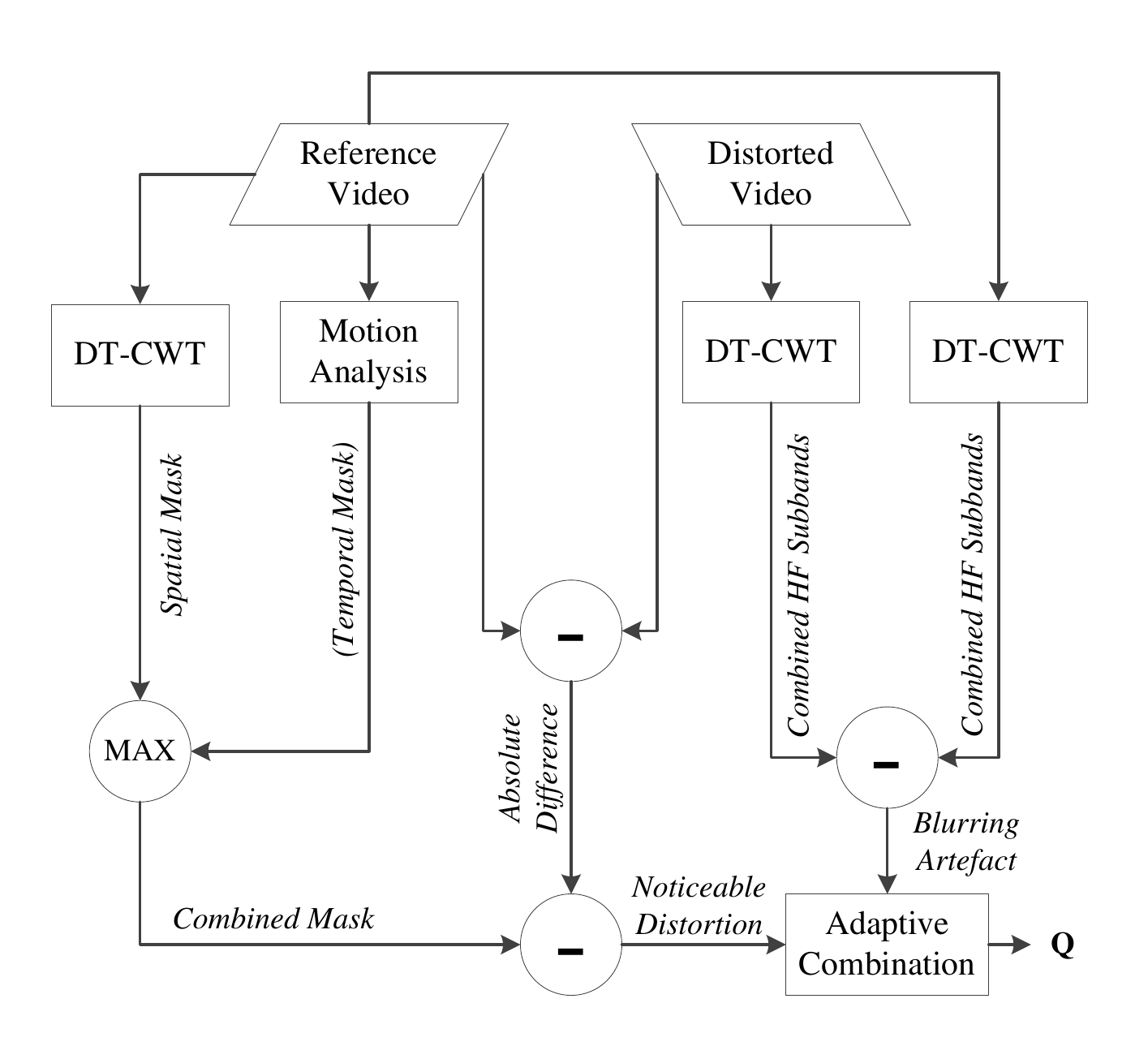}}
  \vspace{-0.1cm}
\caption{The proposed quality assessment method.}

\label{fig:framework}
\end{figure}

The architecture of the perception-based video metric (PVM) is shown in Figure \ref{fig:framework}. This approach is an enhanced version of AVM \cite{j:Zhang} that combines noticeable distortions and blurring artefacts using a modified geometric mean model \cite{j:MAD}. This emulates the distortion-artefact perception process. 

\subsection{Estimating Noticeable Distortion}

The dual-tree complex wavelet transform (DT-CWT) \cite{j:Kingsbury} possesses the important properties of near shift invariance and orientation selectivity and is used here as means of characterising spatial textures. The maximum of all six high frequency DT-CWT subband coefficient magnitudes at the first level of decomposition are used to form a spatial mask.
\begin{equation}
 	M_s(x,y) = \mathrm{max}\left\{ \left| B^o_i(x,y) \right|, i=1,2,\cdots,6\right\}. 
\label{eq:sMask}
\end{equation}

A temporal mask is used to characterise motion regularity and is defined by (\ref{eq:tMask}).
\begin{equation}
M_t(x,y) = \left| \mathrm{SD}(x,y)_x \right| +\left| \mathrm{SD}(x,y)_y \right|.
\label{eq:tMask}
\end{equation}
Here, $\mathrm{SD}(x,y)_x$ and $\mathrm{SD}(x,y)_y$ are the approximated second derivatives of the motion vector $\mathrm{MV}(x,y)$, which are based on an $8\times8$ block at coordinate $(x,y)$ at the block level.

Spatial and temporal masks are merged using equation (\ref{eq:mask}), where $\rho_1$ and $\rho_2$ are combination parameters. These are obtained empirically to achieve the optimum correlation performance based on the VQEG database.
\begin{equation}
	M(x,y) = \mathrm{max}\{\rho_1 \cdot M_{s}(x,y), \rho_2 \cdot M_{t}(x,y)\}.
\label{eq:mask}
\end{equation}

It should be noted that the temporal mask could be omitted, as given by (\ref{eq:maskI}), yielding an approach based only on the current frame. This version requires no motion estimation and is therefore more suitable for in-loop processing. 
\begin{equation}
	M(x,y) = \rho_1 \cdot M_{s}(x,y).
\label{eq:maskI}
\end{equation}

The combined mask is used as a tolerance map to stimulate texture masking effects by thresholding the absolute difference ($\mathrm{AD}$) from the distorted and original frames. This process is described by equation (\ref{eq:AD}) - (\ref{eq:framedifference}).

\begin{equation}
	\mathrm{AD}(x,y) = \left|I_o(x,y) - I_d(x,y)\right|,
\label{eq:AD}
\end{equation}
\begin{equation}
{\mathrm{ND}}(x,y) = {\mathrm{AD}}(x,y) - M(x,y),
\label{eq:pixeldifference1}
\end{equation}
\begin{equation}
{\mathrm{ND}}(x,y) = 
\left\{
\begin{array}{ll}
0, & \mathrm{if} \ {\mathrm{ND}}(x,y) < 0\\
\mathrm{ND}(x,y),& \mathrm{if} \ \mathrm{ND}(x,y) \geq 0
\end{array}
\right.,
\label{eq:pixeldifference2}
\end{equation}
\begin{equation}
\mathcal{D} = \frac{1}{H \cdot W}\sum_{x=1}^{W} {\sum_{y=1}^{H}{ \mathrm{ND}^2(x,y)}},
\label{eq:framedifference}
\end{equation}
Here $I_o(x,y)$ and $I_d(x,y)$ are the luminance values for the original and distorted frames respectively for pixel $(x,y)$. $\mathrm{ND}(x,y)$ is the noticeable distortion, while $\mathcal{D}$ is the mean of squared noticeable distortions at the frame level (frame size is $H \times W$). 

\subsection{Detecting Blurring Artefacts}

\begin{figure*}[ht]
\centering
\begin{minipage}[c]{0.19\linewidth}
  \centering
  \centerline{\includegraphics[width=3.5cm]{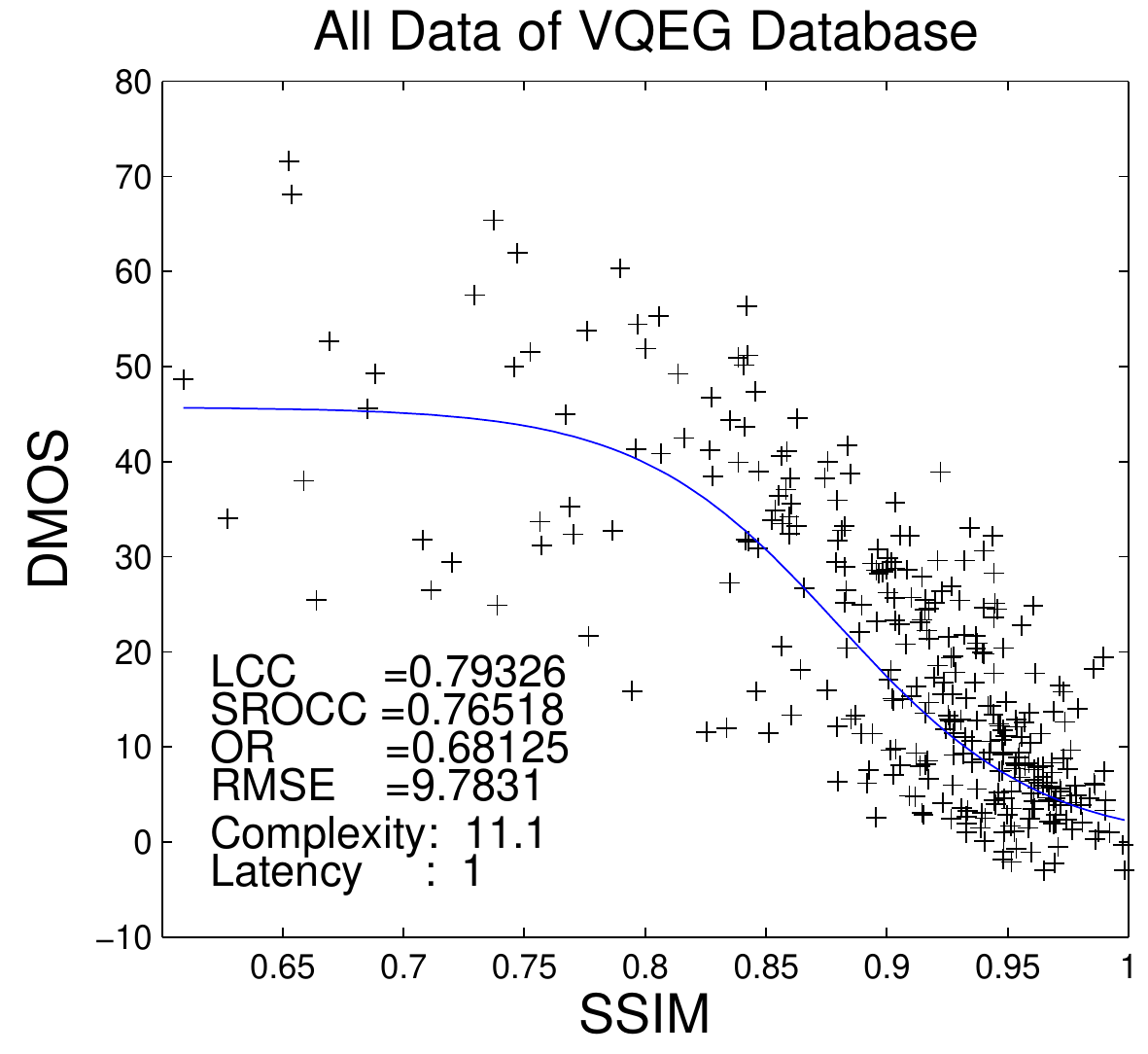}}
  \centerline{(a)}\medskip
\end{minipage}
\begin{minipage}[c]{0.19\linewidth}
  \centering
  \centerline{\includegraphics[width=3.5cm]{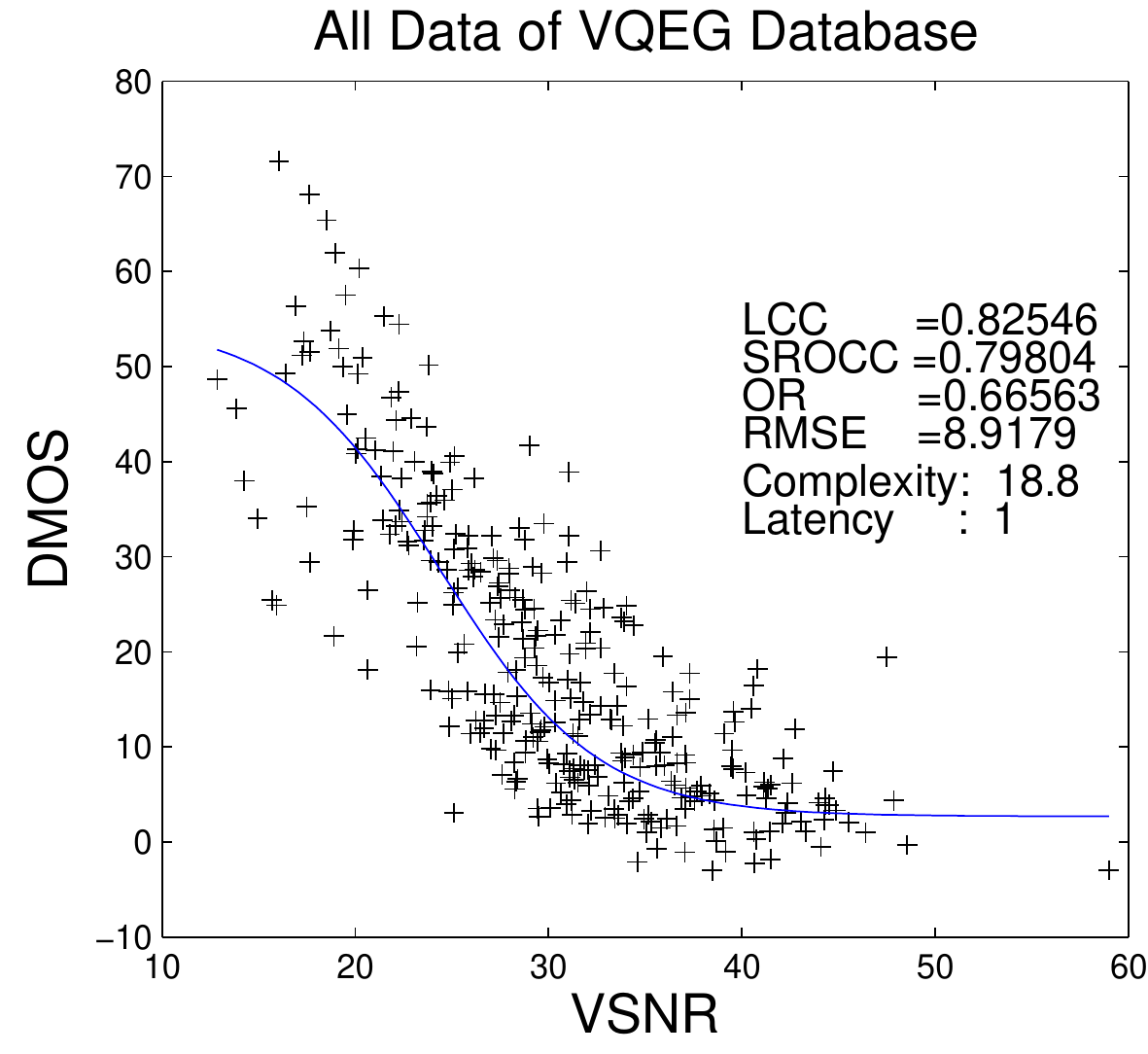}}
  \centerline{(b)}\medskip
\end{minipage}
\begin{minipage}[c]{0.19\linewidth}
  \centering
  \centerline{\includegraphics[width=3.5cm]{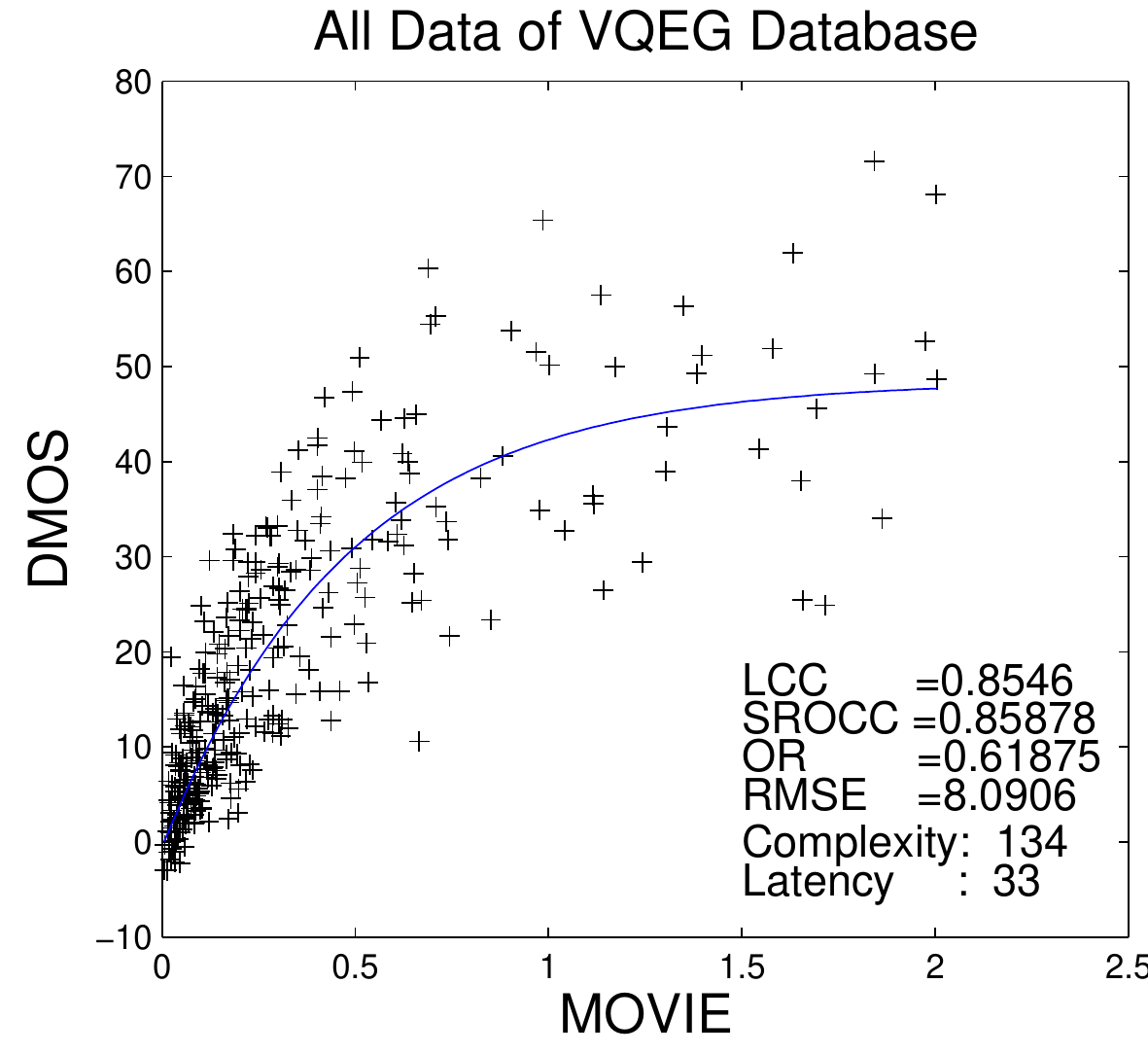}}
  \centerline{(c)}\medskip
\end{minipage}
\begin{minipage}[c]{0.19\linewidth}
  \centering
  \centerline{\includegraphics[width=3.5cm]{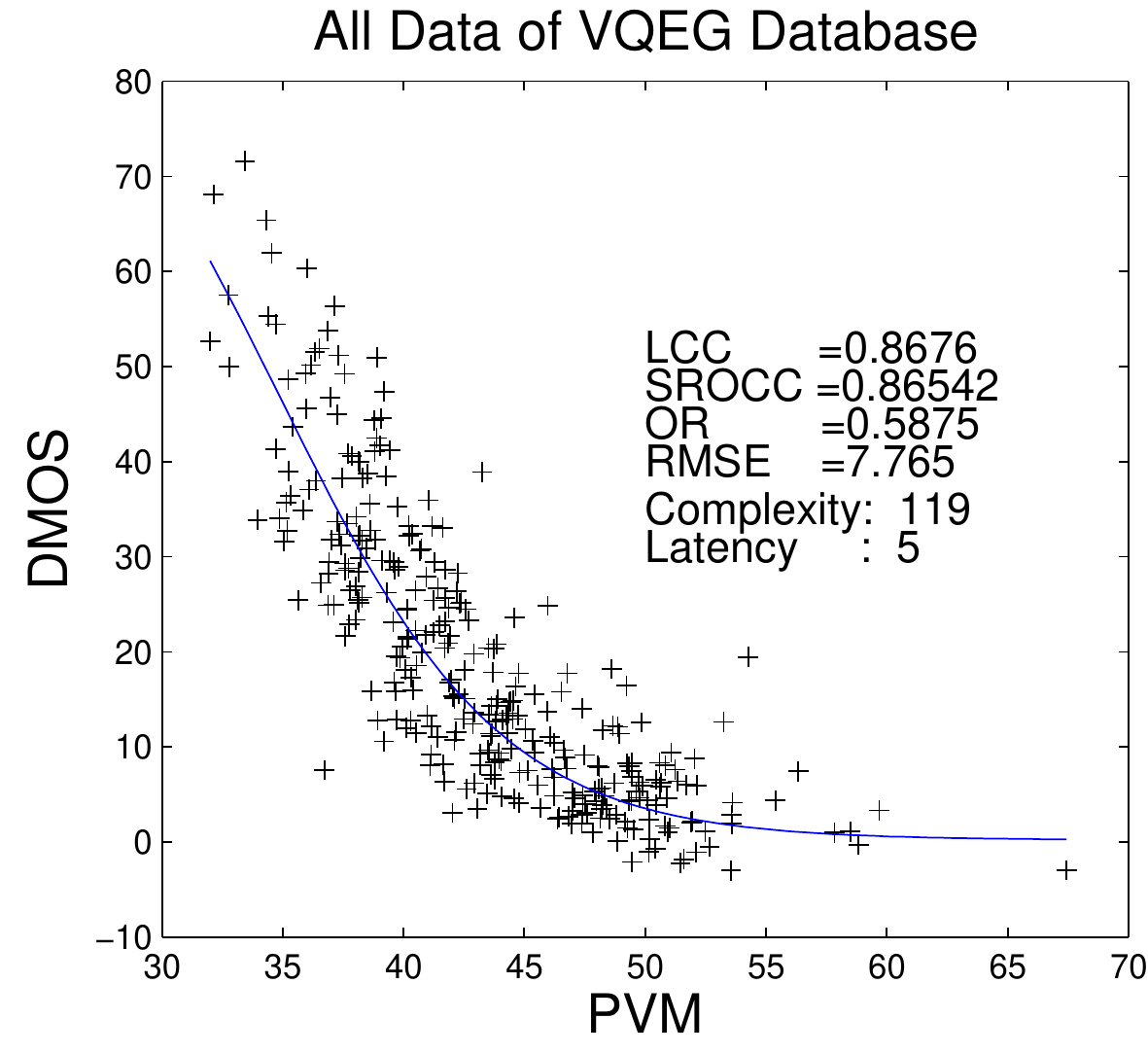}}
  \centerline{(d)}\medskip
\end{minipage}
\begin{minipage}[c]{0.19\linewidth}
  \centering
  \centerline{\includegraphics[width=3.5cm]{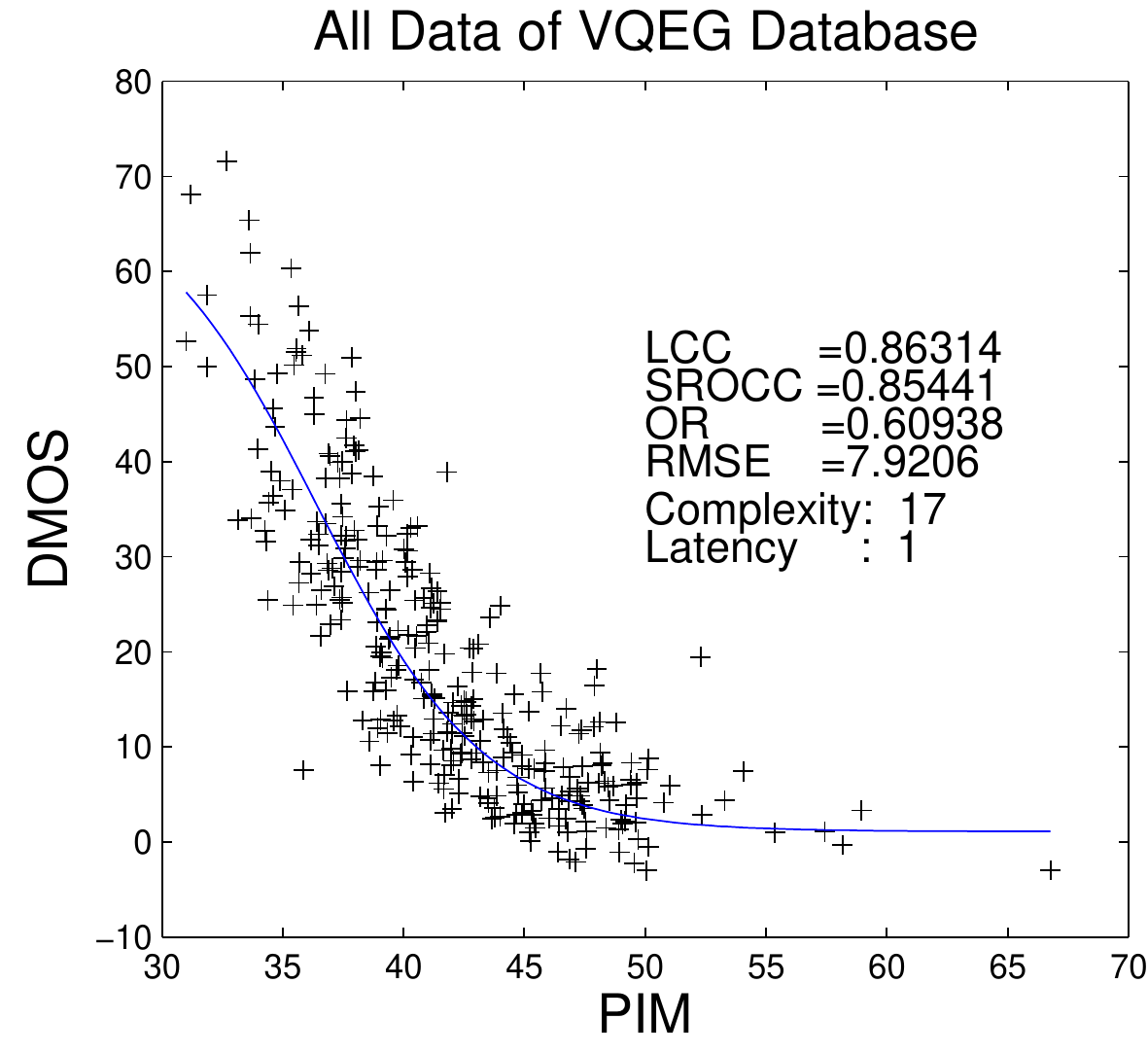}}
  \centerline{(e)}\medskip
\end{minipage}
\caption{Scatter plots of subjective DMOS versus quality metrics on the VQEG database. (a) SSIM. (b) VSNR. (c) MOVIE. (d) PVM. (e) PIM.}
\label{fig:plots}
\end{figure*}

\begin{table*}[ht] 
\caption{Summary of all performance results for the tested metrics.}
		\footnotesize
\centering 
		\begin{tabular}{ccccccc}
\toprule Metric					&   LCC		&		SROCC 	&  OR 		&  	RMSE  			&	Complexity & Latency\\ 
\midrule PSNR 					&  0.7821	&  	0.7832 &  0.6719	&		9.8784 			&	1						& 1\\ 
\midrule SSIM 			 		&  0.7933	&  	0.7652 &  0.6813	&		9.7831 			& 11.1				& 1\\ 
\midrule VSNR 			 		&  0.8255	&  	0.7980 &  0.6656	&		8.9179 			& 18.8				& 1\\ 
\midrule MOVIE 					&  0.8546	&  	0.8588 &  0.6188	&		8.0906 			& 134					&	33\\ 
\midrule PVM 						&  \textbf{0.8676}	&  	\textbf{0.8654} &  \textbf{0.5875}	&		\textbf{7.7650} & 119 (102 for ME)& 5\\
\midrule PIM 						&  0.8631	&  	0.8544 &  0.6094	&		7.9206 			& 17				& 1\\  
\bottomrule 
\end{tabular} 
\label{tab:params}
\end{table*} 

In highly distorted content, especially for compressed material, blurring - a reduction of high frequency energy - is one of the most common artefacts. PVM again uses the DT-CWT \cite{j:Kingsbury} to compare high frequency subband coefficients between the original and distorted frames, averaging the difference to obtain the frame level blurring ($\mathcal{B}$), as given in equations (\ref{eq:blurring1}) - (\ref{eq:frameblurring}).
\begin{equation}
\mathrm{BL}(x, y)= \sum^{6}_{i=1} {| B^o_i(x,y) |} - \sum^{6}_{i=1} {| B^d_i(x,y) |},
\label{eq:blurring1}
\end{equation}
\begin{equation}
\mathrm{BL}(x, y)=  \left\{
\begin{array}{ll}
0, & \mathrm{if} \  \mathrm{BL}(x,y)< 0 \\
\mathrm{BL}(x, y), &\mathrm{if}\  \mathrm{BL}(x,y)\geq 0
\end{array}
\right.,
\label{eq:blurring2}
\end{equation}
\begin{equation}
\mathcal{B} = \frac{1}{H \cdot W}\sum_{x=1}^{W} {\sum_{y=1}^{H}{ \mathrm{BL}(x,y)}}.
\label{eq:frameblurring}
\end{equation}%
Here  $|B^d_i(x,y)|$ is the amplitude of one of the six subbands coefficients after decomposition at pixel $(x,y)$ within the distorted frame. 

\subsection{Adaptive Pooling}

According to the distortion-artefact perception strategy, we employ a weighted geometric mean to obtain the frame-level quality index between two videos. This has been previously used in \cite{j:MAD}, and is given by equations (\ref{eq:model1}) and (\ref{eq:model2}).

\begin{equation}
Q = \mathcal{D}^{\alpha}\cdot (\chi \mathcal{B})^{1-\alpha},
\label{eq:model1}
\end{equation}
where $\alpha$ is obtained by:
\begin{equation}
\alpha = \frac{1}{1 + \beta_1 \mathcal{D}^{\beta_2}}.
\label{eq:model2}
\end{equation}
Here $\beta_1$, $\beta_2$ and $\chi$ are pre-determined parameters based on the VQEG database. It can be observed that an additional parameter $\chi$ is added compared to the method in \cite{j:MAD} to improve the freedom of the model.

Finally, a video quality index is obtained by averaging quality indices for all frames ($\bar{Q}$) and converting to decibel units ($Q_{dB}$).

\begin{equation}
Q_{dB} = 10 \cdot log(\frac{255^2}{\bar{Q}}).
\label{eq:model3}
\end{equation}

\section{Results and Evaluation}
\label{sec:results}
\begin{figure*}[ht]
\centering
\begin{minipage}[c]{0.19\linewidth}
  \centering
  \centerline{\includegraphics[width=3.5cm]{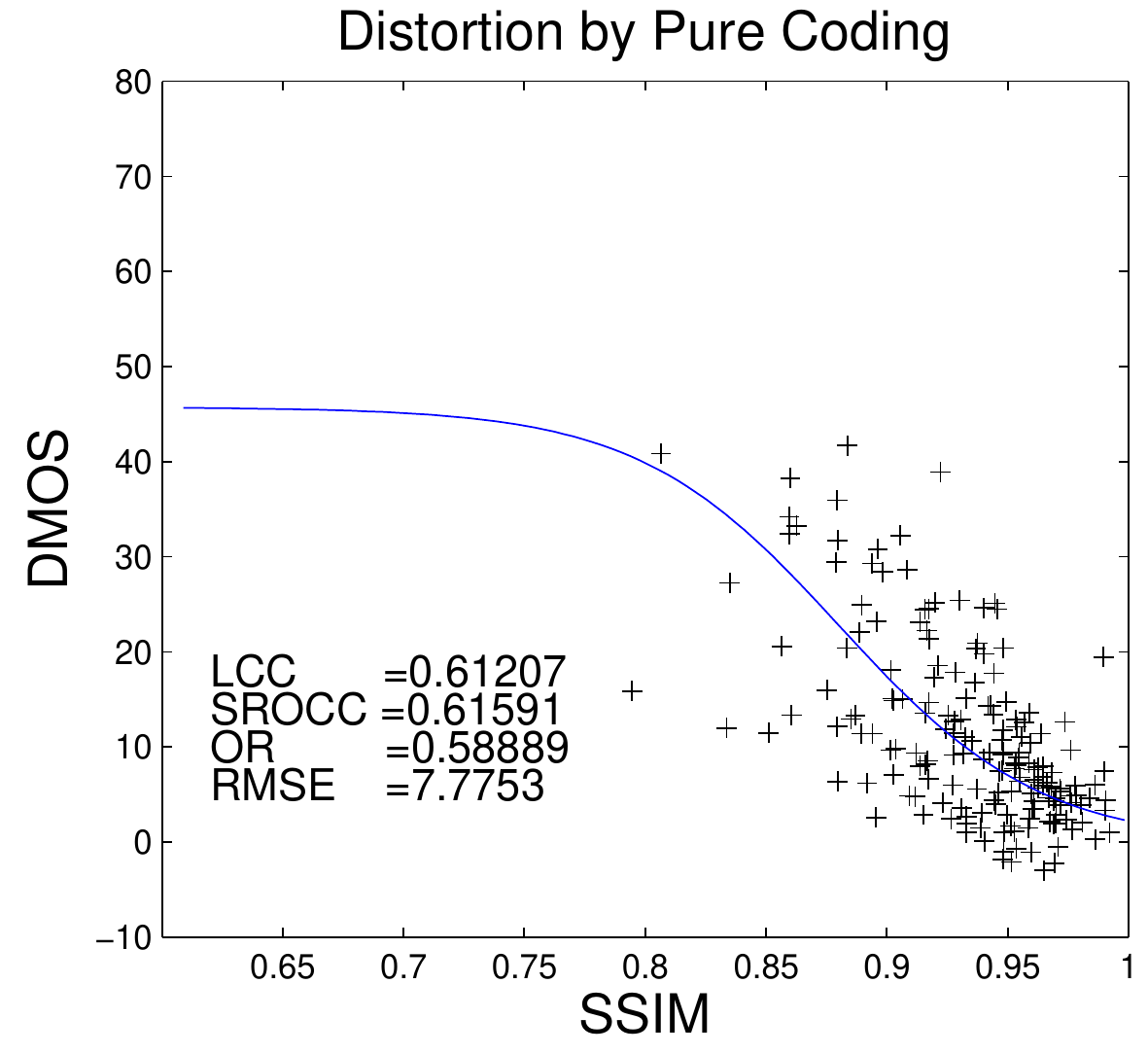}}
  \centerline{(a)}\medskip
\end{minipage}
\begin{minipage}[c]{0.19\linewidth}
  \centering
  \centerline{\includegraphics[width=3.5cm]{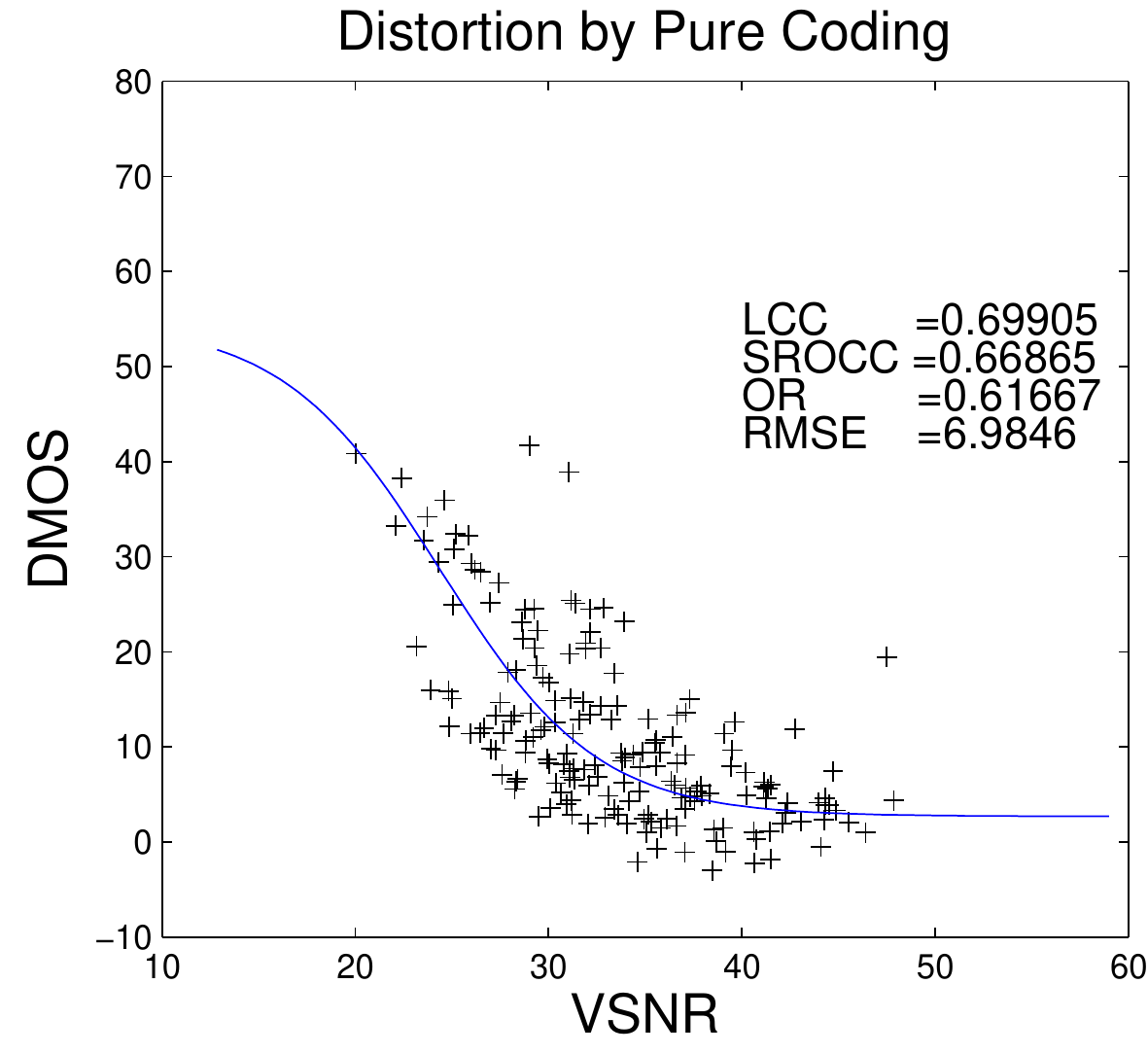}}
  \centerline{(b)}\medskip
\end{minipage}
\begin{minipage}[c]{0.19\linewidth}
  \centering
  \centerline{\includegraphics[width=3.5cm]{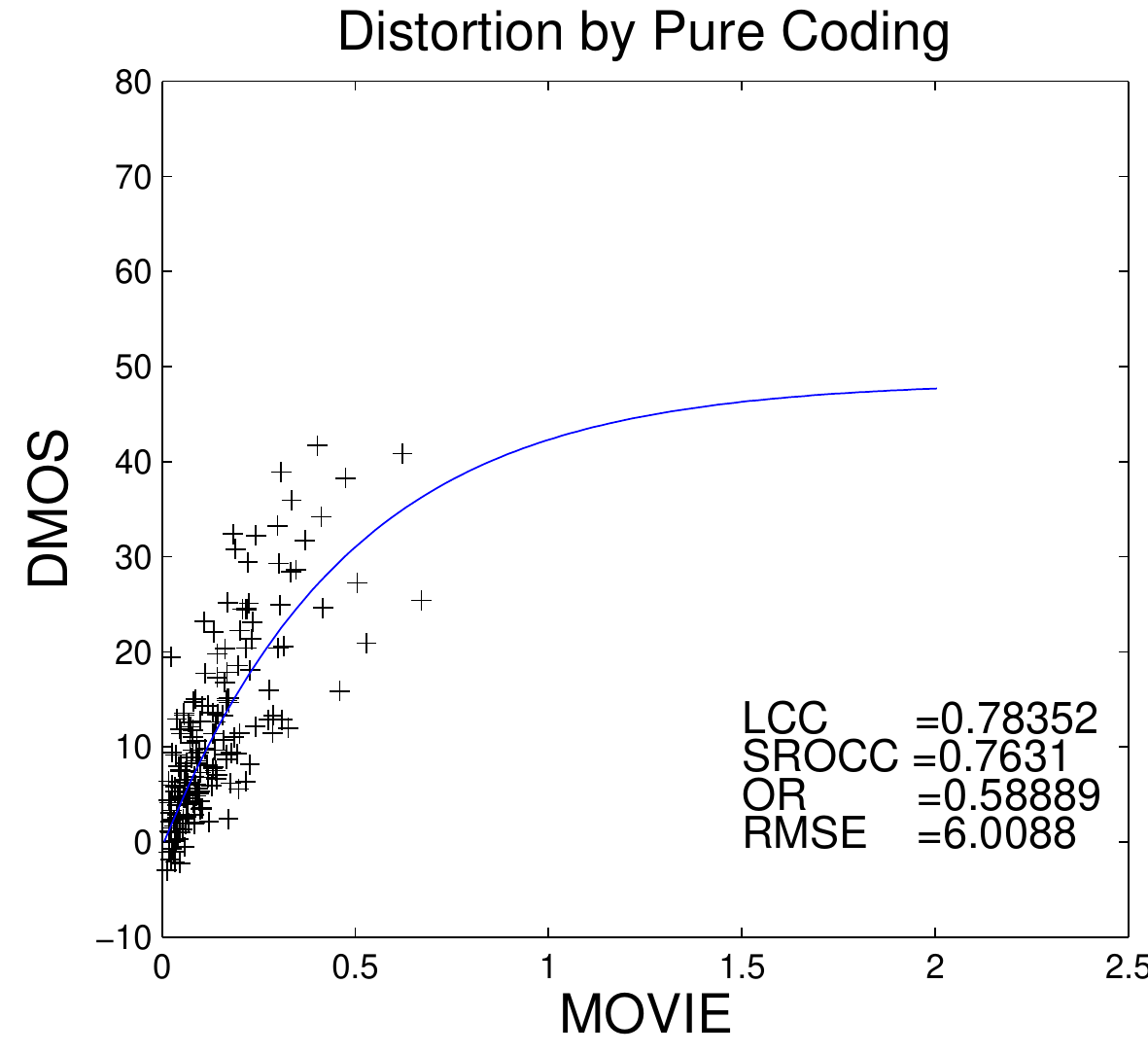}}
  \centerline{(c)}\medskip
\end{minipage}
\begin{minipage}[c]{0.19\linewidth}
  \centering
  \centerline{\includegraphics[width=3.5cm]{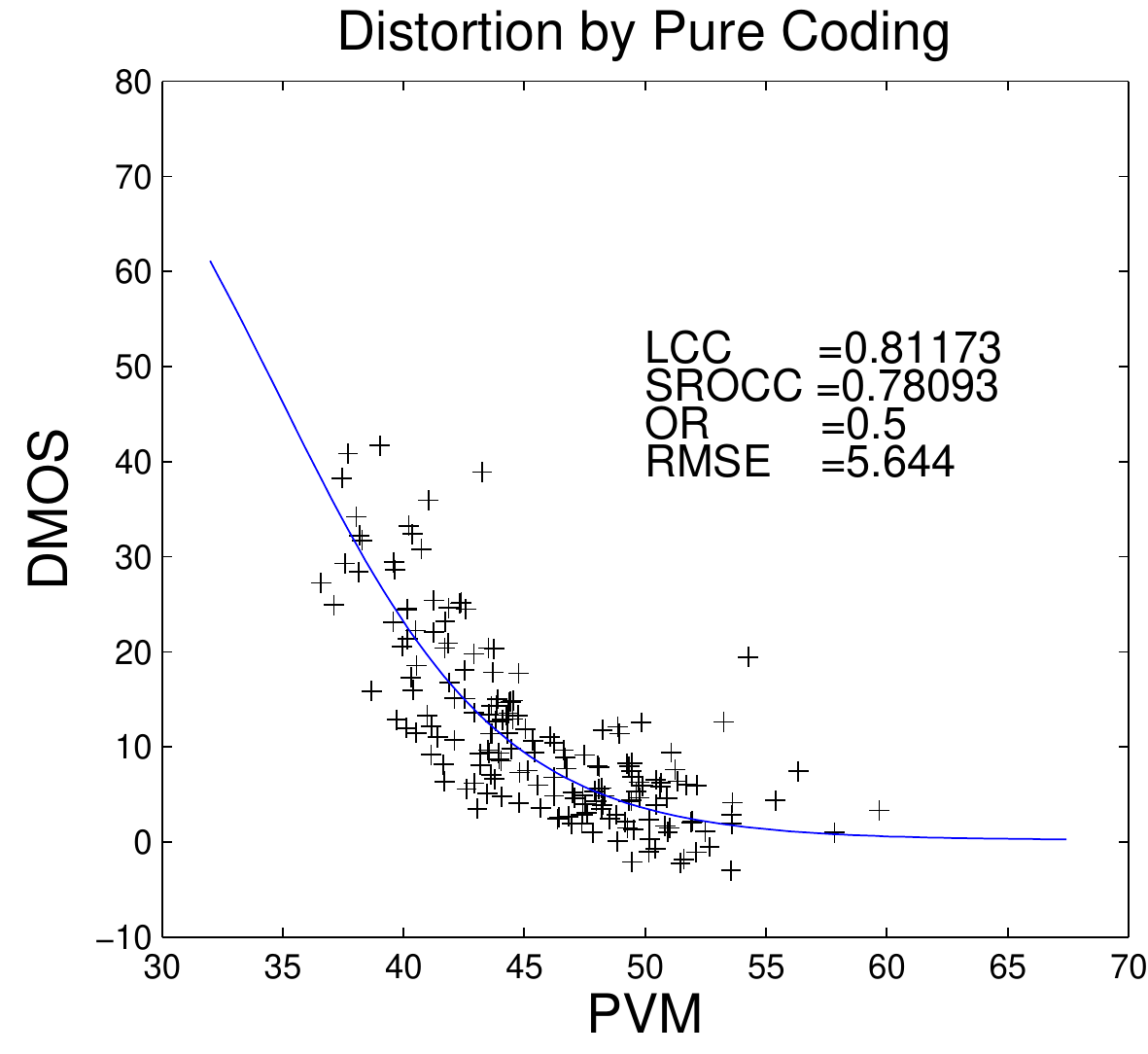}}
  \centerline{(d)}\medskip
\end{minipage}
\begin{minipage}[c]{0.19\linewidth}
  \centering
  \centerline{\includegraphics[width=3.5cm]{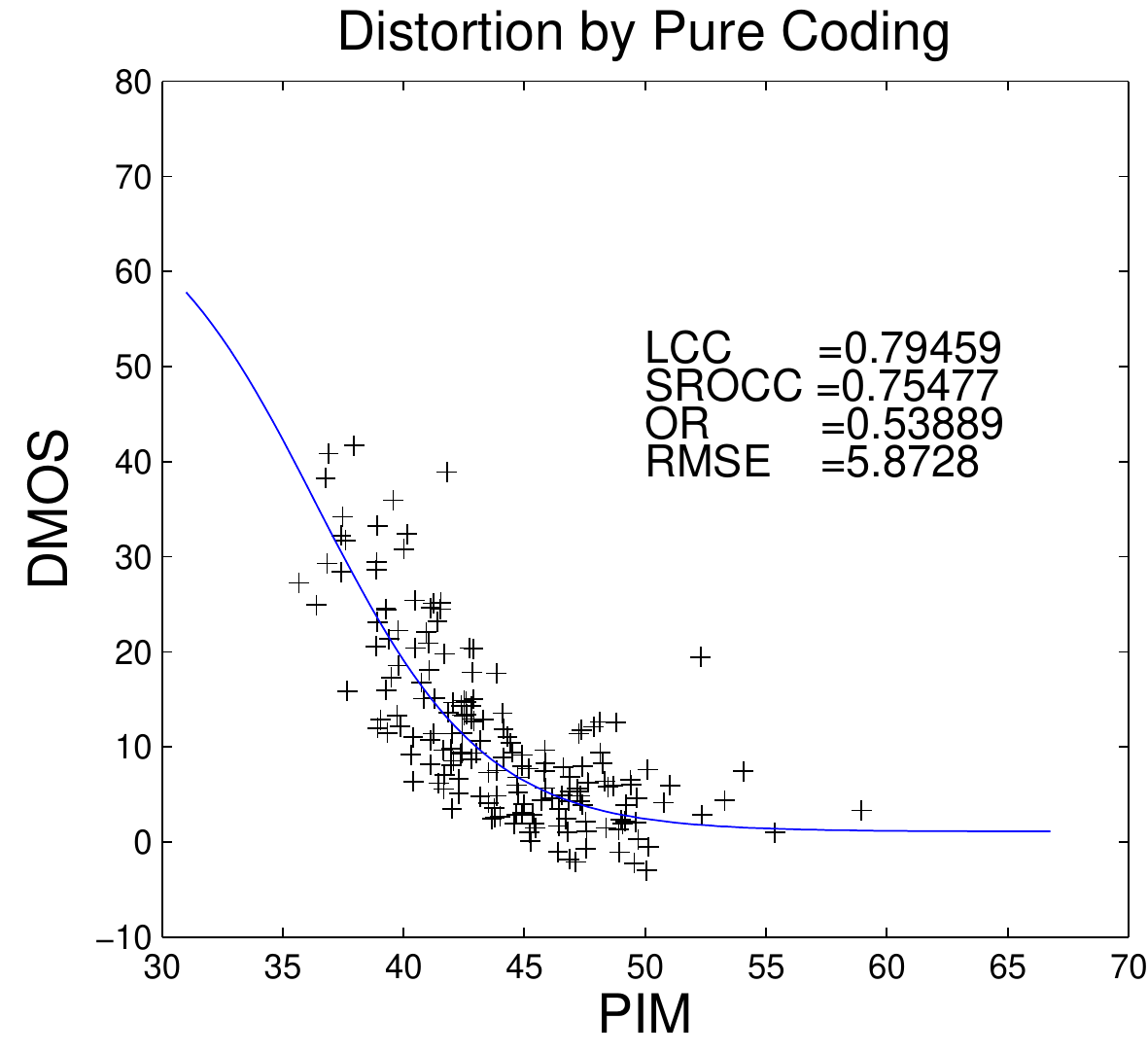}}
  \centerline{(e)}\medskip
\end{minipage}
\begin{minipage}[c]{0.19\linewidth}
  \centering
  \centerline{\includegraphics[width=3.5cm]{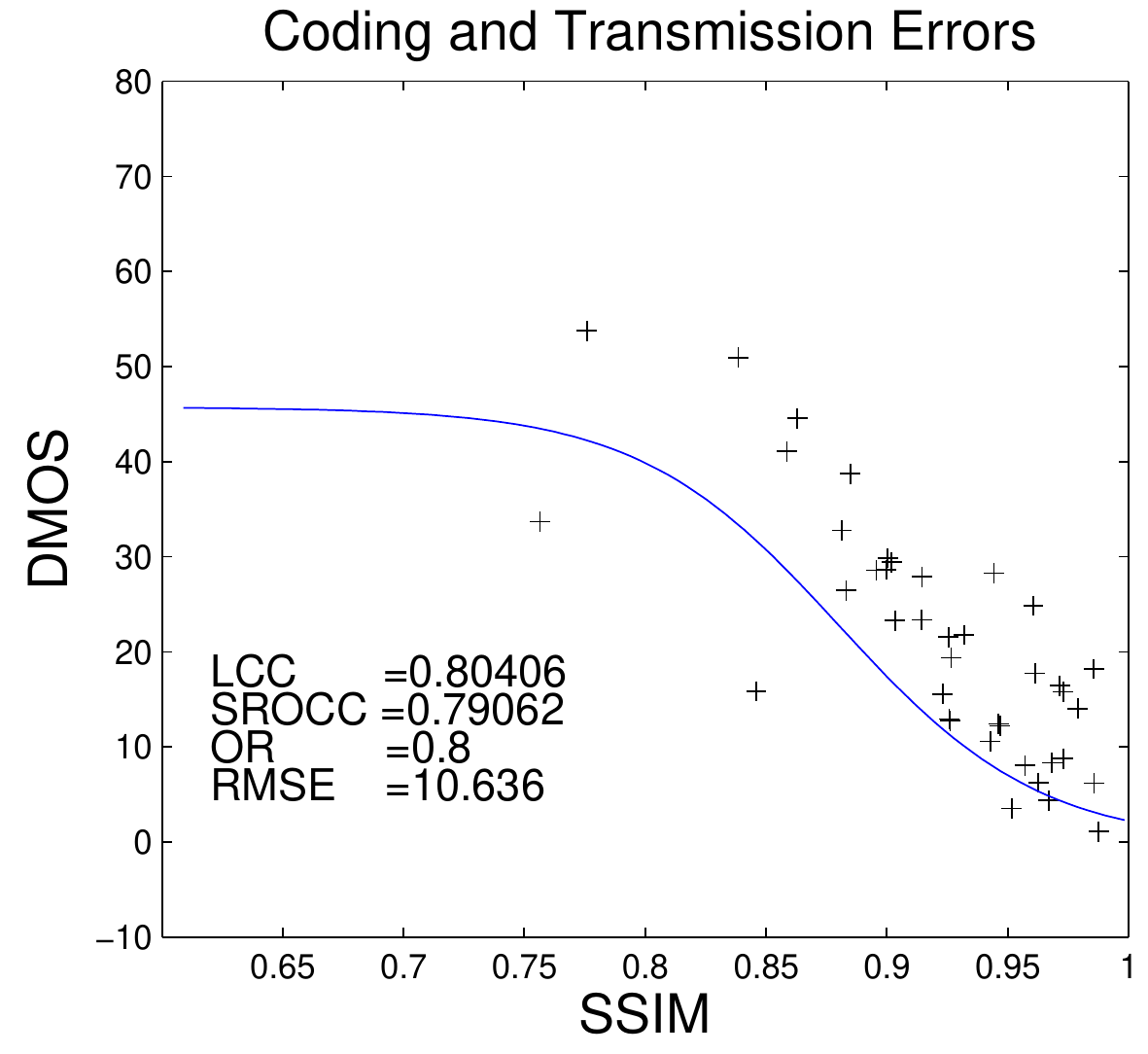}}
  \centerline{(f)}\medskip
\end{minipage}
\begin{minipage}[c]{0.19\linewidth}
  \centering
  \centerline{\includegraphics[width=3.5cm]{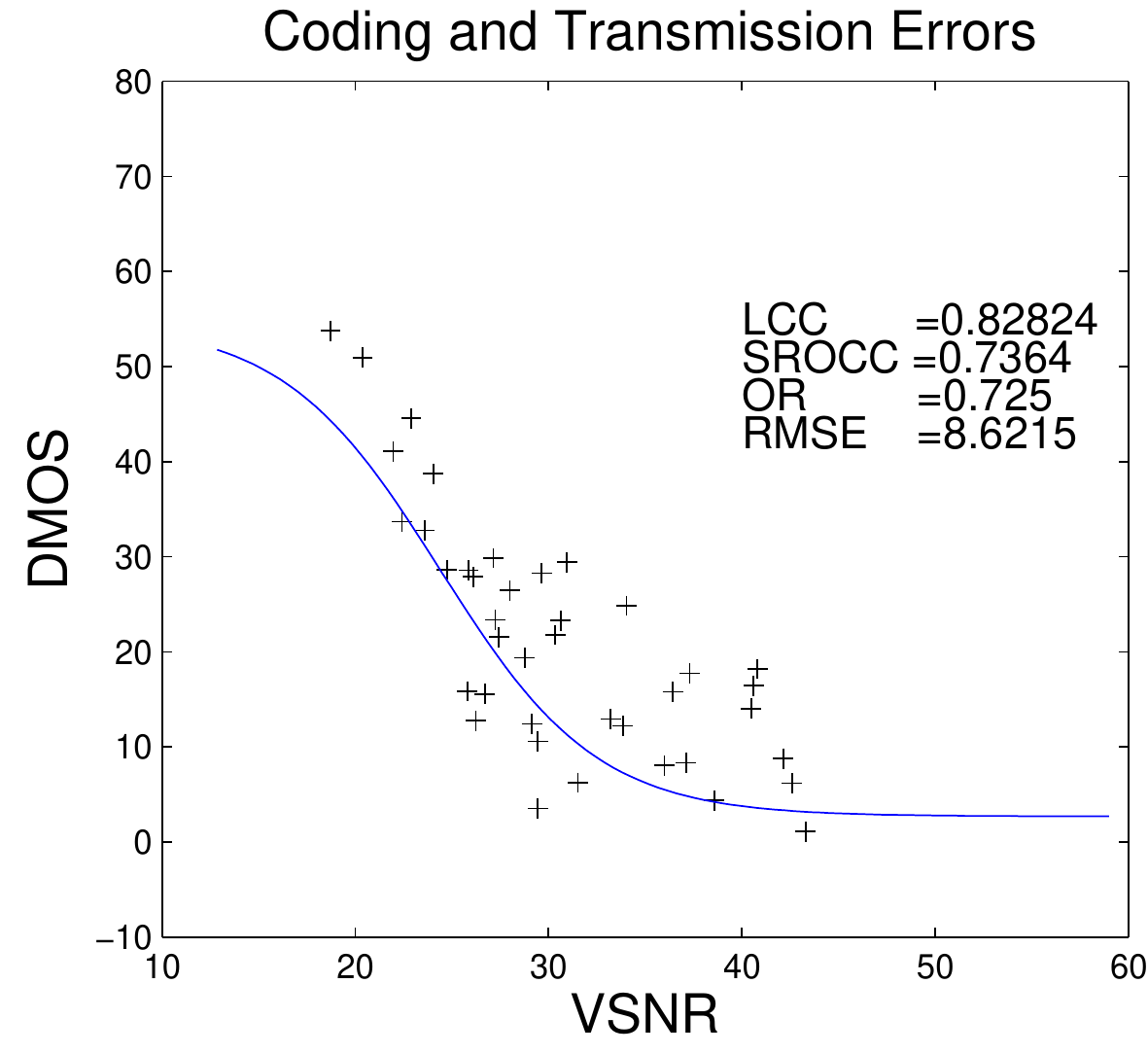}}
  \centerline{(g)}\medskip
\end{minipage}
\begin{minipage}[c]{0.19\linewidth}
  \centering
  \centerline{\includegraphics[width=3.5cm]{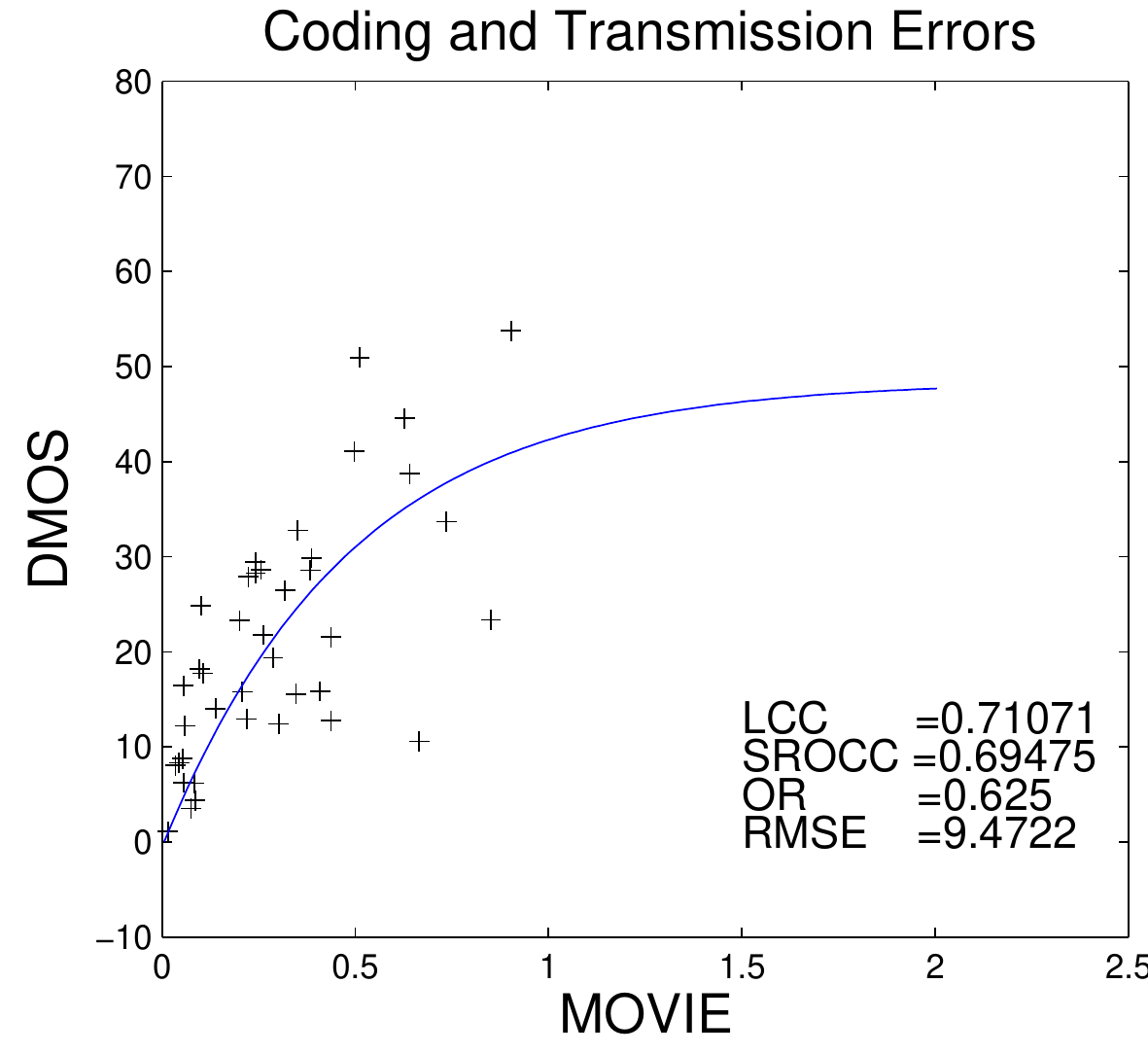}}
  \centerline{(h)}\medskip
\end{minipage}
\begin{minipage}[c]{0.19\linewidth}
  \centering
  \centerline{\includegraphics[width=3.5cm]{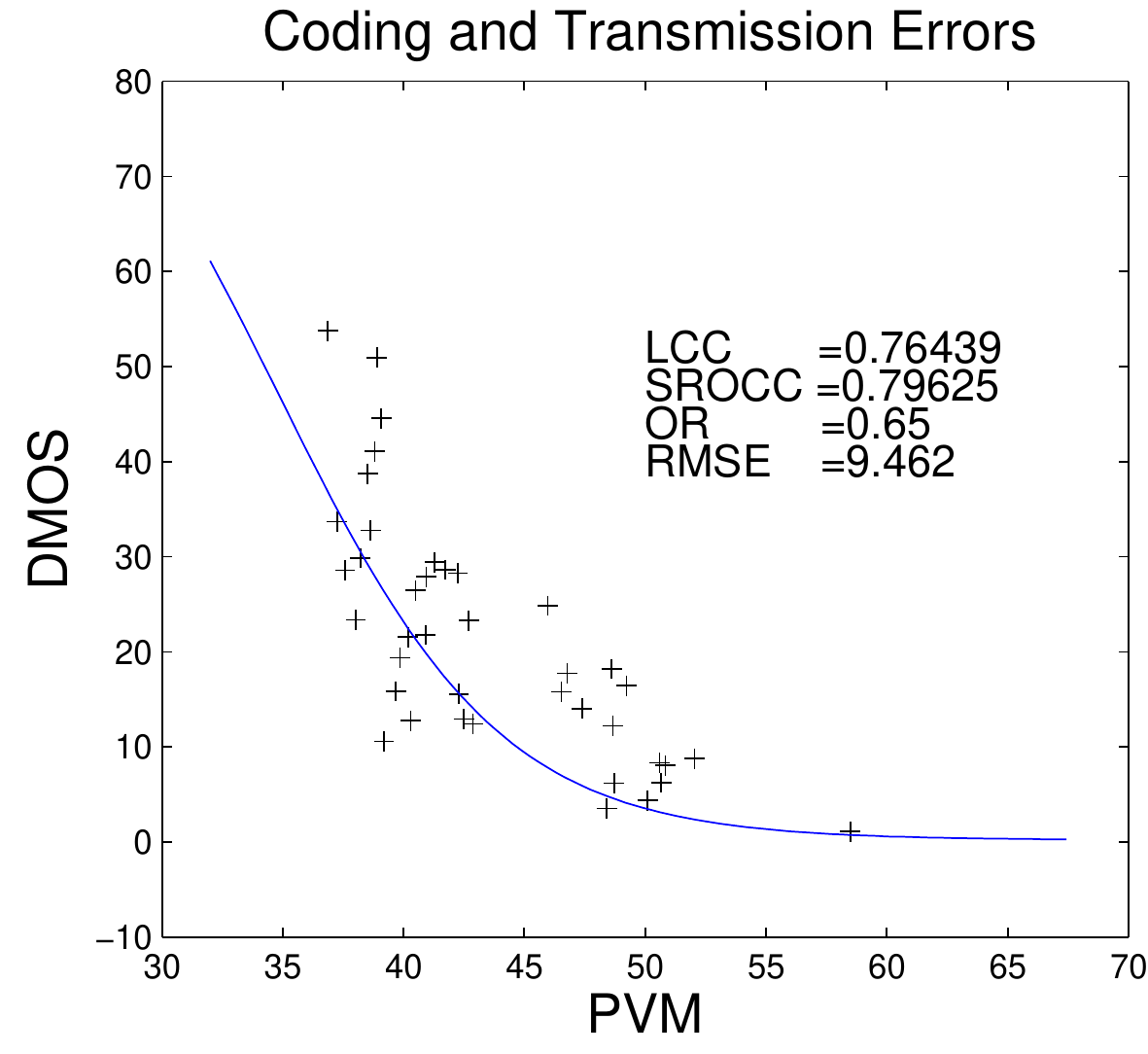}}
  \centerline{(i)}\medskip
\end{minipage}
\begin{minipage}[c]{0.19\linewidth}
  \centering
  \centerline{\includegraphics[width=3.5cm]{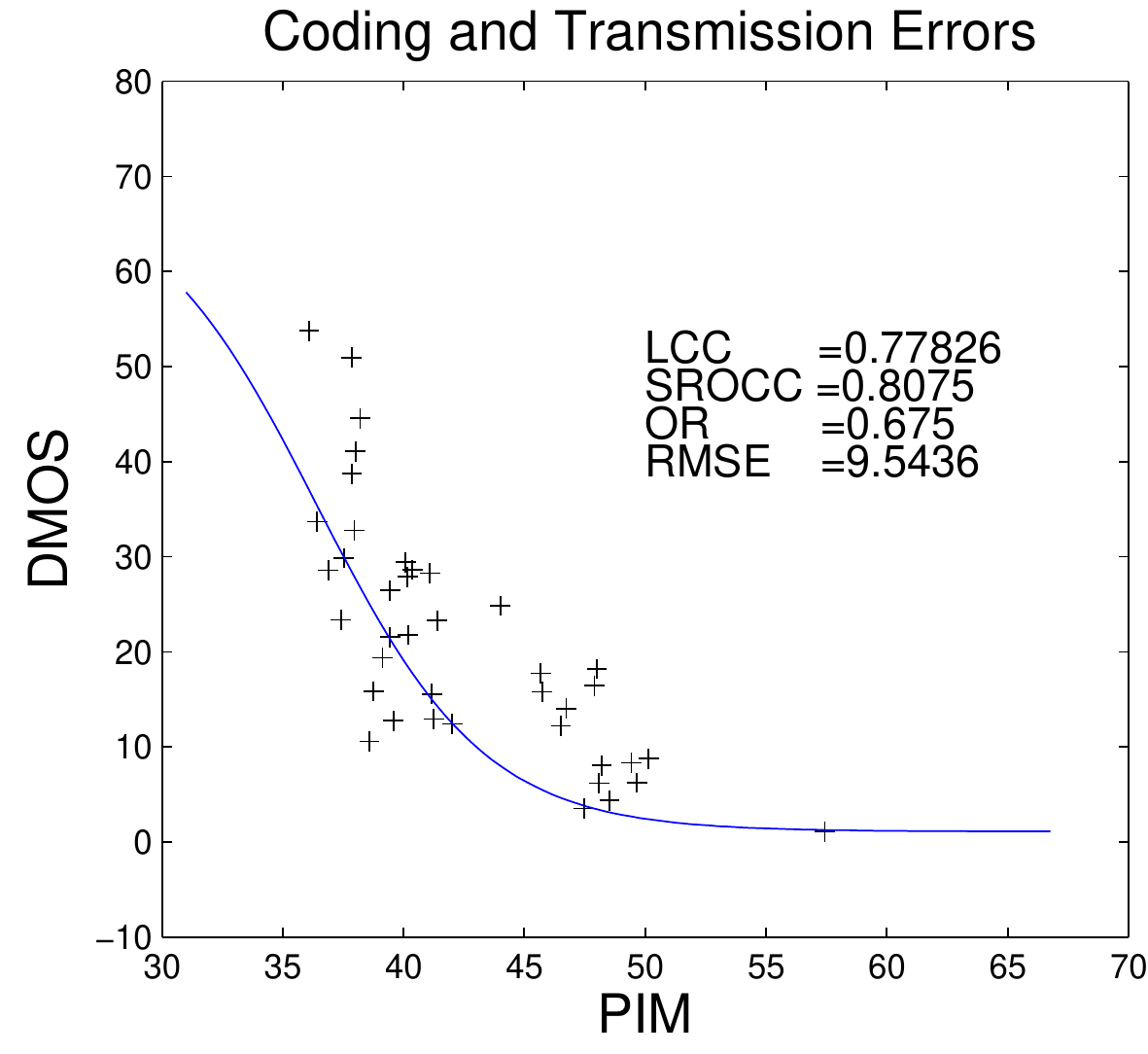}}
  \centerline{(j)}\medskip
\end{minipage}
\begin{minipage}[c]{0.19\linewidth}
  \centering
  \centerline{\includegraphics[width=3.5cm]{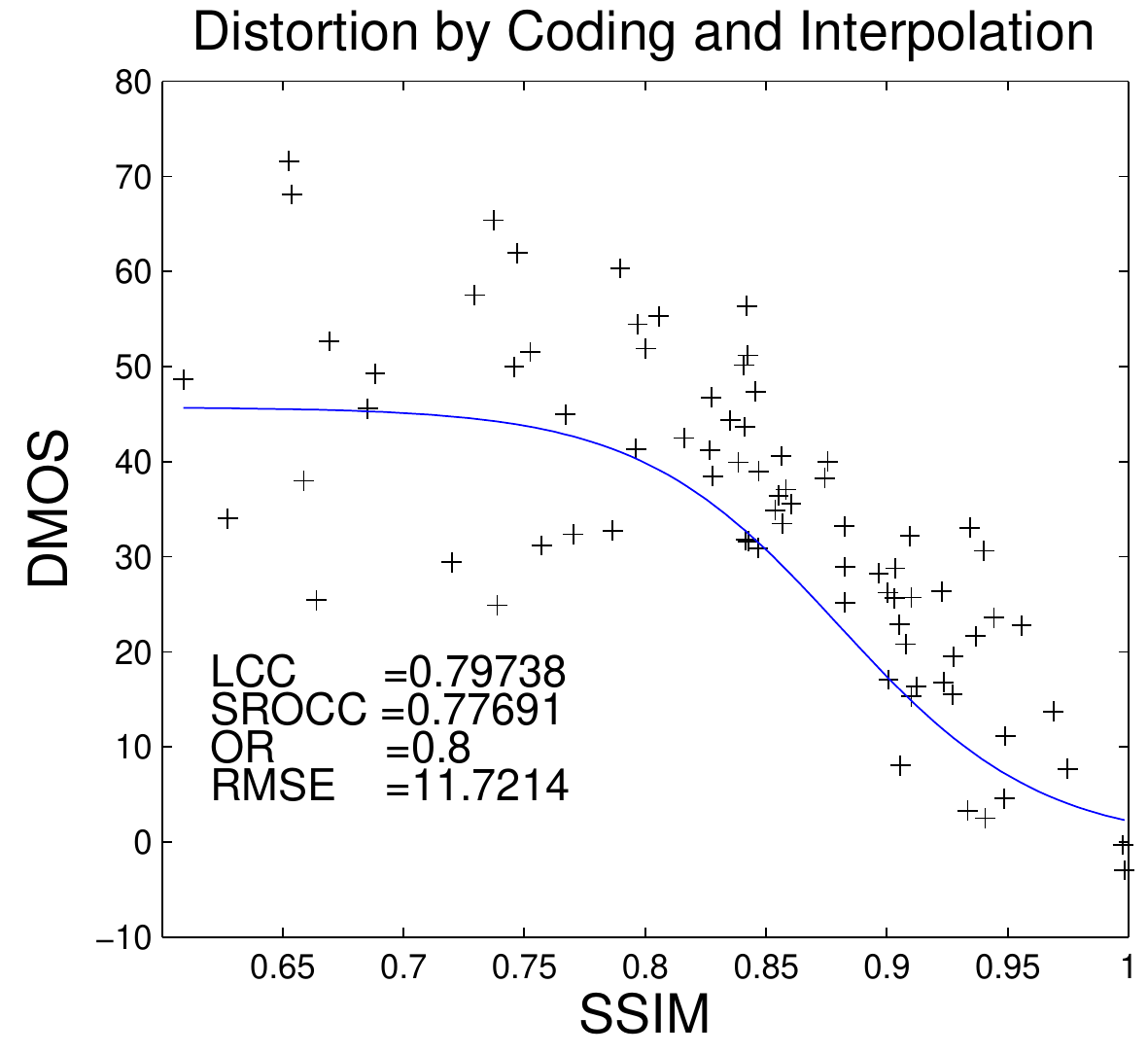}}
  \centerline{(k)}\medskip
\end{minipage}
\begin{minipage}[c]{0.19\linewidth}
  \centering
  \centerline{\includegraphics[width=3.5cm]{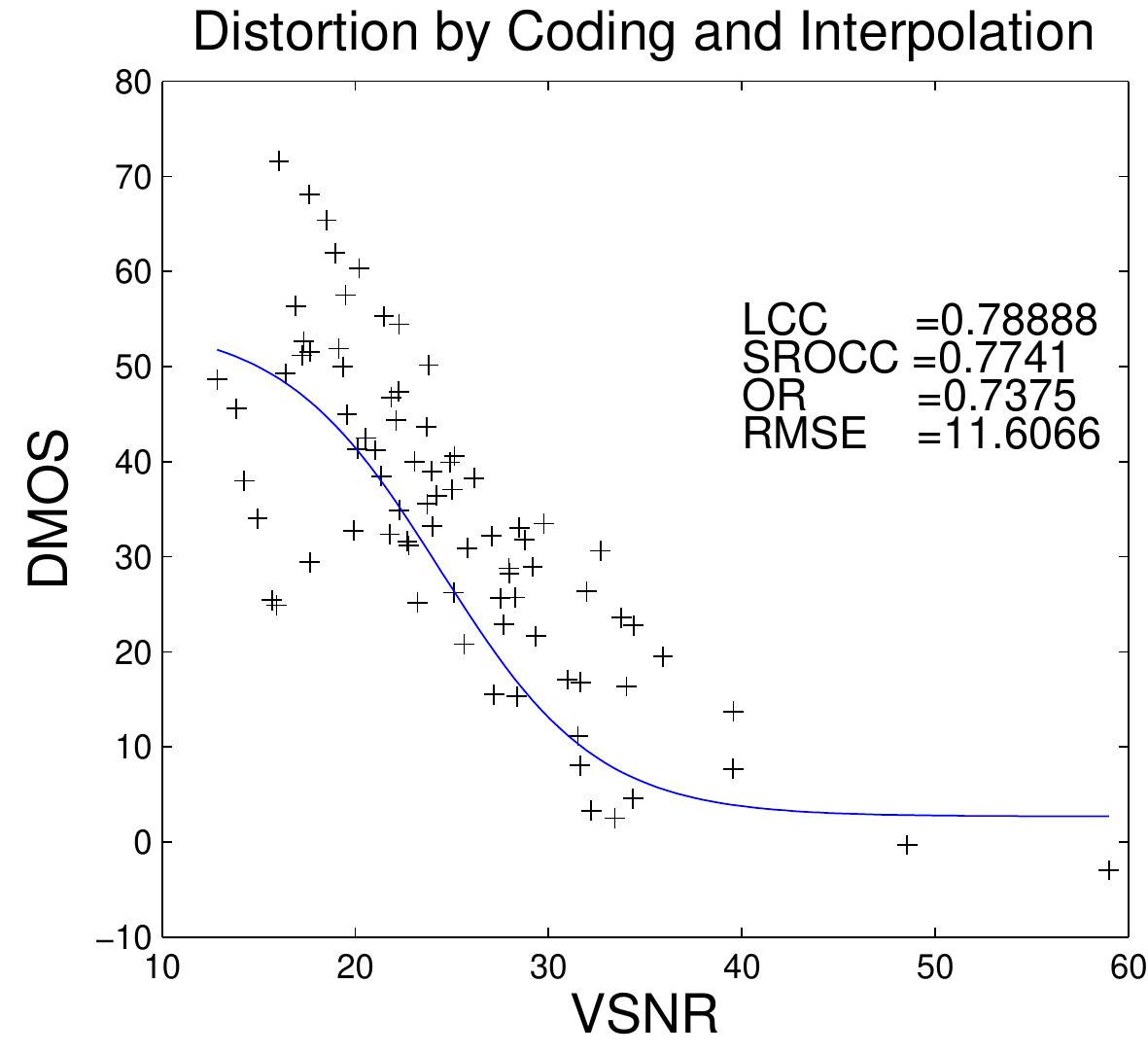}}
  \centerline{(l)}\medskip
\end{minipage}
\begin{minipage}[c]{0.19\linewidth}
  \centering
  \centerline{\includegraphics[width=3.5cm]{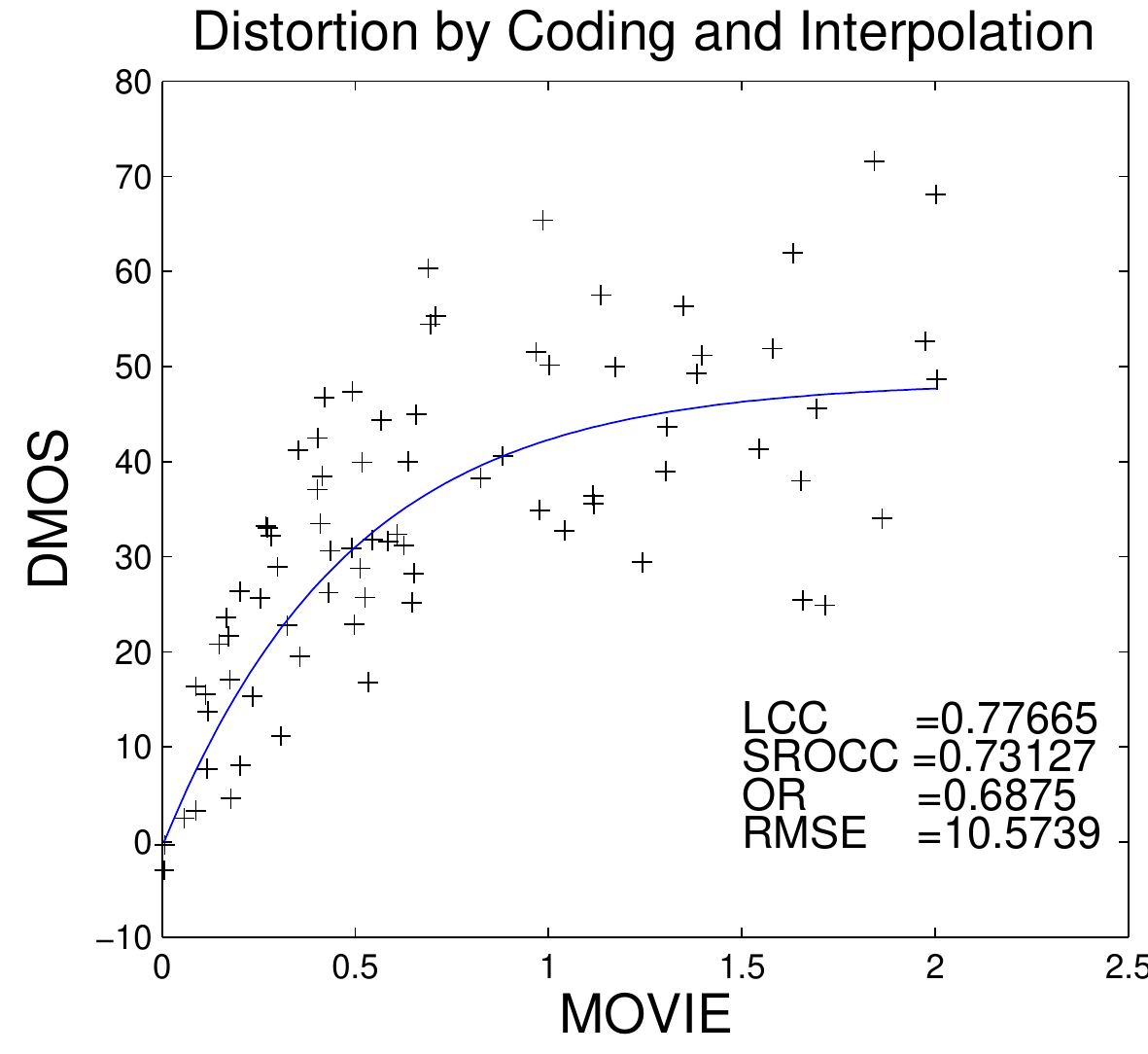}}
  \centerline{(m)}\medskip
\end{minipage}
\begin{minipage}[c]{0.19\linewidth}
  \centering
  \centerline{\includegraphics[width=3.5cm]{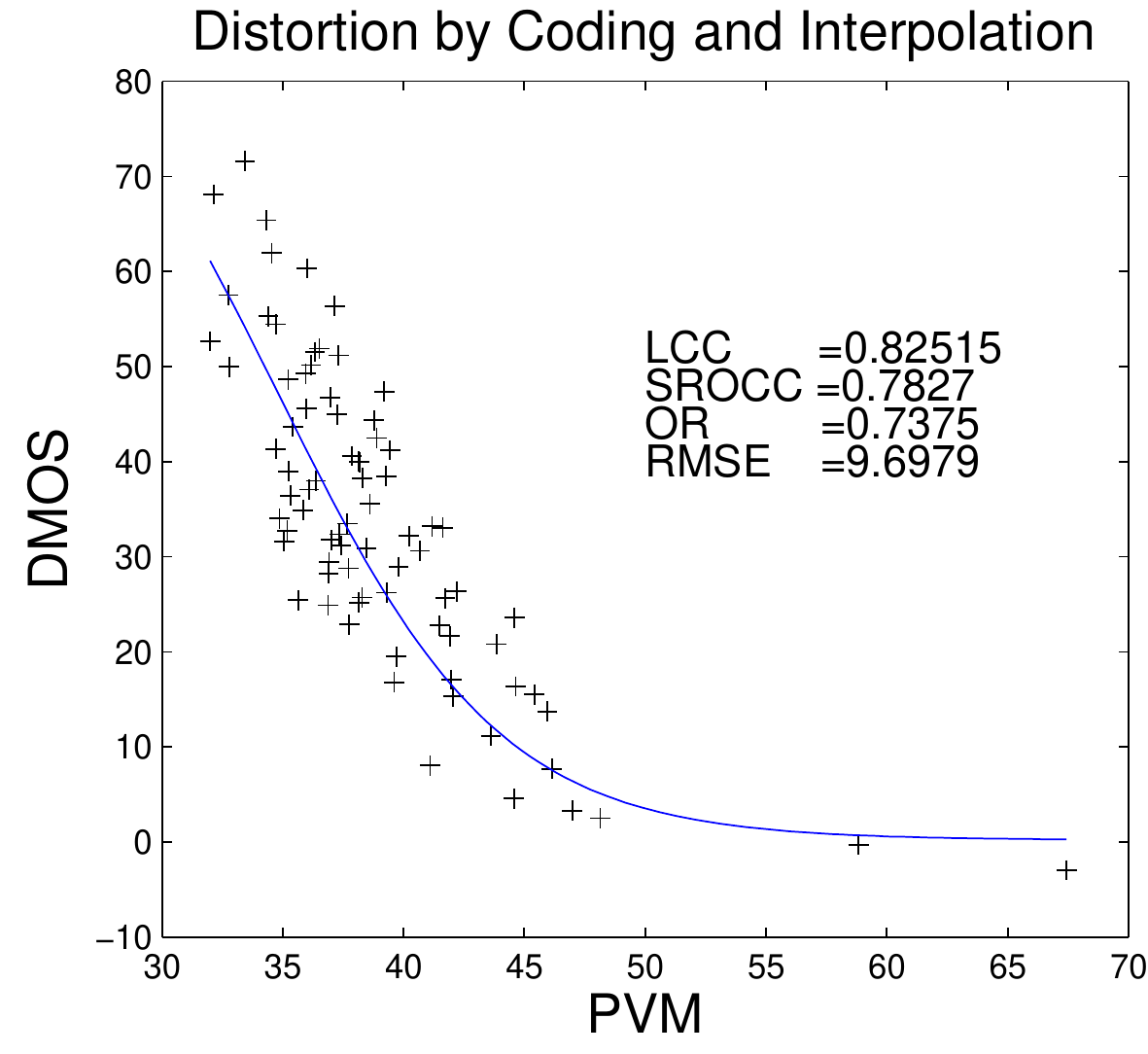}}
  \centerline{(n)}\medskip
\end{minipage}
\begin{minipage}[c]{0.19\linewidth}
  \centering
  \centerline{\includegraphics[width=3.5cm]{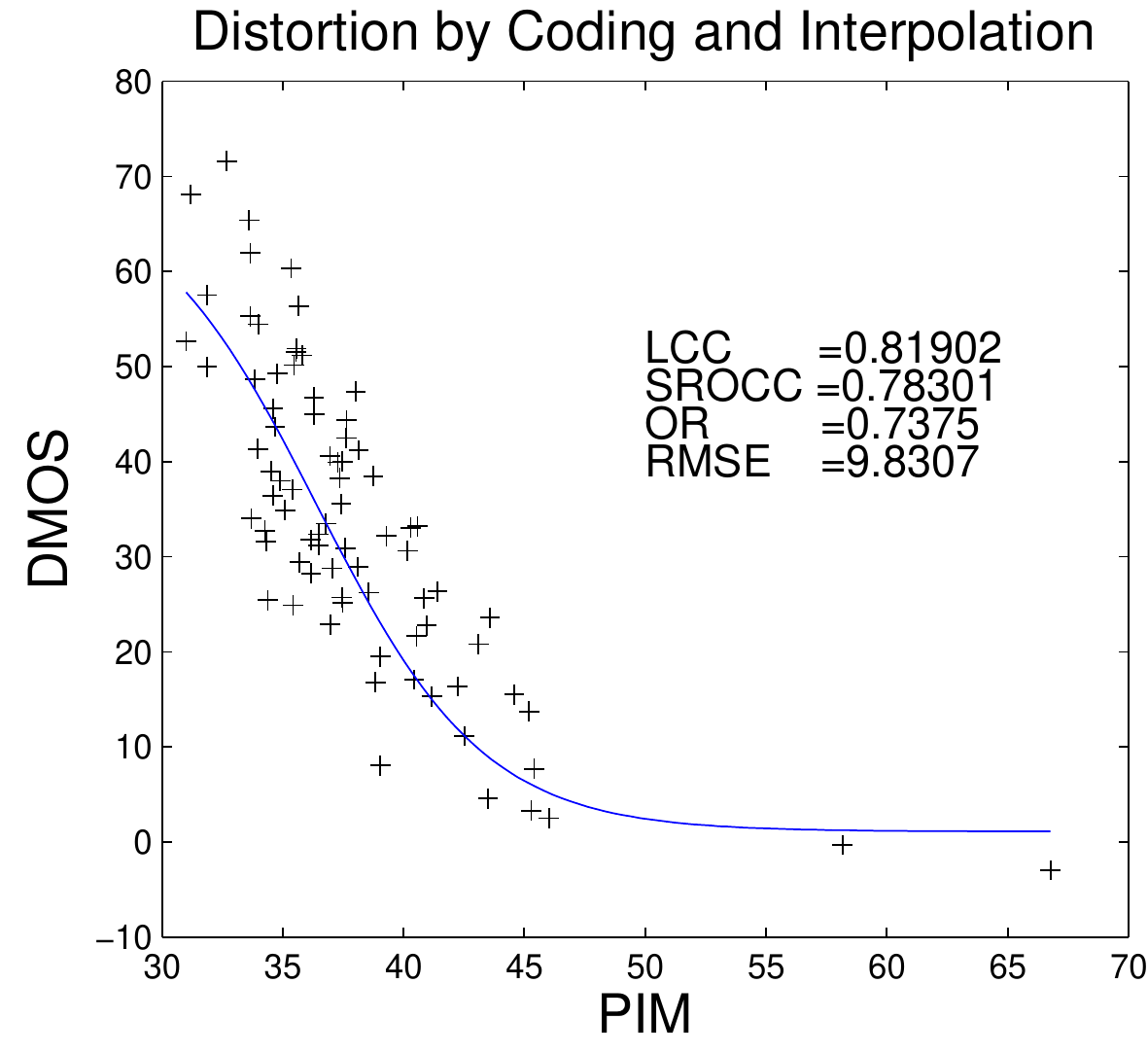}}
  \centerline{(o)}\medskip
\end{minipage}
\caption{Scatter plots of subjective DMOS versus quality metrics for sub groups of the VQEG database. (a-e) Distortion by conventional compression for SSIM, VSNR, MOVIE, PVM and PIM respectively. (f-j) Distortion by compression and artificial errors for these metrics. (k-o) Distortion by compression and interpolation for these metrics.}
\label{fig:plots1}
\end{figure*}

For in-loop mode selection and RQO applications in compression, the candidate video metric should provide accurate prediction of subjective opinions, manageable computational complexity and low latency. These three factors are all analysed below.

In order to characterise the ability of  predicting subjective video quality, the proposed perception-based video metric (PVM) and its lite version (by disregarding the temporal mask), PIM, were fully evaluated using the VQEG FRTV Phase 1 database \cite{r:vqeg}, along with the analysis rules used in \cite{r:vqeg1}. A weighted least squares approach is utilised to obtain the logistic fitting function \cite{r:vqeg}.  To keep consistency with other researchers \cite{j:movie}, for distortion types HRC 8 and 9 (there are two different subjective scores), only ``low quality'' DMOS is used. All sequences were cropped by 20 pixels in all four directions in order to deal with the overscan display effect. SSIM \cite{j:ssim}, VSNR \cite{j:vsnr} and MOVIE \cite{j:movie} are used as references. MOVIE is generally considered as the state of the art in objective quality metrics, providing precise subjective quality prediction for the VQEG database. SSIM and VSNR both offer lower computational complexity and latency, and are therefore more flexible for integration into a compression framework.

The correlation performance can be demonstrated by scatter point plots, and the results for these metrics are shown in Figure \ref{fig:plots}. The scatter plots for PVM are much more compact around the fitting curve than those for SSIM, VSNR and MOVIE across the whole quality range. An approximately linear relationship exists between PVM (or PIM) and the subjective scores, and this is valuable for in-loop RQO integration. This performance can be further analysed by dividing the whole database into three sub groups based on distortion types: compressed videos, compressed content with transmission errors, and compressed and interpolated materials, as shown in Figure \ref{fig:plots1}.

After non-linear regression, statistical parameters include the Linear Correlation Coefficient (LCC), the Spearman Rank Order Correlation Coefficient (SROCC), the Outlier Ratio (OR) and the Root Mean Squared Error (RMSE) are obtained to evaluate the correlation performance of these metrics. These values are shown in Figure \ref{fig:plots}. Definitions of these four parameters can be found in \cite{r:vqeg,j:movie}. PVM is superior to SSIM, VSNR and MOVIE, showing higher correlation, fewer outliers and lower squared errors. This improvement is also evident for its lite version - PIM. 

From the results in Figure \ref{fig:plots1}, it is observed that PVM and PIM provide excellent correlation across all distortion types. With the exception of transmission errors, where VSNR, MOVIE and PVM all exhibit good performance, PVM provides the best performance.

In the case of perceptual video coding, most distortion types arise from compression and texture warping and synthesis involving spatiotemporal interpolation \cite{c:NdjikiNya,j:Zhang}. For this type of distortion (pure coding and coding with interpolation) within the VQEG database, PVM and PIM both perform better than other tested metrics. 

The computational complexity and latency of each quality metric is also recorded on the graphs in Figure \ref{fig:plots}. The values for complexity are normalised based on the execution time of PSNR. The relative execution time for PVM is much longer than that for PSNR, SSIM and VSNR, and is only slightly less than MOVIE. It should be noted however that motion estimation is a dominant factor for PVM, and for practical in-loop applications, this may reuse data produced during the coding process. In this case, the relative complexity for PVM can be significantly reduced from 119 to 17, which is similar to that of PIM. The latency for all metrics is defined as the number of frames used for computing the quality index of a frame. Conventionally MOVIE uses 33 frames for computing one frame level quality index, while only five frames are used in PVM. This can be further reduced to 1 by applying its lite version - PIM. A summary of all performance results is given in Table \ref{tab:params}.

\section{Conclusions}
\label{sec:conclusion}

This paper has characterised  the performance of new perception-based video quality metrics (PVM and PIM) on a range of representative distortion types. The proposed PVM and its lite version PIM provide competitive performance in contrast to existing video quality assessment approaches such as SSIM, VSNR, and MOVIE. It also offers the benefit of low complexity and latency and is therefore suitable for in-loop RQO. Future research should focus on the RQO application using the proposed metrics.

\small
\bibliographystyle{IEEEbib}
\bibliography{IEEEabrv,MyRef}

\begin{thebibliography}{10}

\bibitem{j:Lee}
J.S. Lee and T.~Ebrahimi,
\newblock ``Perceptual video compression: a survey,''
\newblock {\em IEEE Journal of Selected Topics in Signal Processing}, vol. 6,
  pp. 684--697, 2012.

\bibitem{j:Kelly}
D.~H. Kelly,
\newblock ``Motion and vision. {II}. stabilized spatio-temporal threshold
  surface,''
\newblock {\em Journal of Optical Society of America}, vol. 69, no. 10, pp.
  1340--1349, 1979.

\bibitem{j:ZhangXue}
X.~Zhang, W.~Lin, and P~Xue,
\newblock ``Improved estimation for just-noticeable visual distortion,''
\newblock {\em Signal Processing}, vol. 84, no. 4, pp. 795--808, 2005.

\bibitem{j:Wei}
Z.~Wei and K.~N. Ngan,
\newblock ``Spatio-temporal just noticeable distortion profile from grey scale
  image/video in dct domain,''
\newblock {\em IEEE Trans. on Circuits and System for Video Technology}, vol.
  19, no. 3, pp. 337--346, 2009.

\bibitem{j:vsnr}
D.~Chandler and S.~Hemami,
\newblock ``{VSNR}: A wavelet-based visual signal-to-noise ratio for natural
  images,''
\newblock {\em IEEE Trans. on Image Processing}, vol. 16, no. 9, pp.
  2284--2298, 2007.

\bibitem{c:NdjikiNya}
P.~Ndjiki-Nya, T.~Hinz, and T.~Wiegand,
\newblock ``Generic and robust video coding with texture analysis and
  synthesis,''
\newblock in {\em Proc. IEEE Int Conf. on Multimedia \& Expo}. IEEE, 2007, pp.
  1447--1450.

\bibitem{j:Bosch}
M.~Bosch, F.~Zhu, and E.~J. Delp,
\newblock ``Segmentation-based video compression using texture and motion
  models,''
\newblock {\em IEEE Journal of Selected Topics in Signal Processing}, vol. 5,
  no. 7, pp. 1277--1281, 2011.

\bibitem{j:Zhang}
F.~Zhang and D.~R. Bull,
\newblock ``A parametric framework for video compression using region-based
  texture models,''
\newblock {\em IEEE Journal of Selected Topics in Signal Processing}, vol. 5,
  no. 7, pp. 1378--1392, 2011.

\bibitem{j:Chikkerur}
S.~Chikkerur, V.~Sundaram, M.~Reisslein, and L.~J. Karam,
\newblock ``Objective video quality assessment methods: a classification,
  review and performance comparison,''
\newblock {\em {IEEE} Trans. Broadcast.}, vol. 30, pp. 17--26, 2005.

\bibitem{j:ssim}
Z.~Wang, A.~Bovik, H.~Sheikh, and E.~Simoncelli,
\newblock ``Image quality assessment: from error visibility to structural
  similarity,''
\newblock {\em IEEE Trans. on Image Processing}, vol. 13, pp. 600--612, 2004.

\bibitem{j:MAD}
E.~C. Larson and D.~M. Chandler,
\newblock ``Most apparent distortion: full-reference image quality assessment
  and the role of strategy,''
\newblock {\em Journal of Electronic Imaging}, vol. 19, no. 1, pp.
  011006(1--21), 2010.

\bibitem{j:Kingsbury}
N.~G. Kingsbury,
\newblock ``Complex wavelets for shift invariant analysis and filtering of
  signals,''
\newblock {\em Journal of Applied and Computational Harmonic Analysis}, vol.
  10, no. 3, pp. 234--253, 2001.

\bibitem{r:vqeg}
{Video Quality Experts Group},
\newblock ``Final report from the video quality experts group on the validation
  of objective quailty metrics for video quality assessment.,''
\newblock Tech. {R}ep., VQEG, 2000.

\bibitem{r:vqeg1}
{ITU-T Study Group 12},
\newblock ``Evaluation of new methods for objective testing of video quality:
  objective test plan,''
\newblock Tech. {R}ep., ITU-T, 1998.

\bibitem{j:movie}
K.~Seshadrinathan and A.~C. Bovik,
\newblock ``Motion tuned spatio-temporal quailty assessment of natural
  videos,''
\newblock {\em IEEE Trans. on Image Processing}, vol. 19, pp. 335--350, 2010.

\end{thebibliography}

\end{document}